\documentstyle[12pt,equations]{article}
\begin{document}
\newcommand{\tcr}{T_{cr}}
\newcommand{\chit}{\tilde{\chi}}
\newcommand{\phit}{\tilde{\phi}}
\newcommand{\df}{\delta \phi}
\newcommand{\dr}{\delta \rho}
\newcommand{\dxl}{\delta x_{\Lambda}}
\newcommand{\dkl}{\delta \kappa_{\Lambda}}
\newcommand{\dk}{\delta \kappa}
\newcommand{\dlt}{\delta \tilde{\lambda}}
\newcommand{\dkg}{\delta \kappa_{G}}
\newcommand{\dkcr}{\delta \kappa_{cr}}
\newcommand{\dxg}{\delta x_{G}}
\newcommand{\dx}{\delta x}
\newcommand{\lx}{\lambda}
\newcommand{\Lx}{\Lambda}
\newcommand{\ex}{\epsilon}
\newcommand{\ks}{k_s}
\newcommand{\gb}{\bar{g}}
\newcommand{\gt}{\tilde{g}}
\newcommand{\lb}{{\bar{\lambda}}}
\newcommand{\lbz}{\bar{\lambda}_0}
\newcommand{\lt}{\tilde{\lambda}}
\newcommand{\lr}{{\lambda}_R}
\newcommand{\xr}{x_R}
\newcommand{\xt}{\tilde{x}}
\newcommand{\lrt}{{\lambda}_R(T)}
\newcommand{\lbr}{{\bar{\lambda}}_R}
\newcommand{\lk}{{\lambda}(k)}
\newcommand{\lbk}{{\bar{\lambda}}(k)}
\newcommand{\lbkt}{{\bar{\lambda}}(k,T)}
\newcommand{\ltk}{\tilde{\lambda}(k)}
\newcommand{\mx}{{m}^2}
\newcommand{\mxk}{{m}^2(k)}
\newcommand{\mxkt}{{m}^2(k,T)}
\newcommand{\mb}{{\bar{m}}}
\newcommand{\mt}{\tilde{m}}
\newcommand{\mr}{{m}^2_R}
\newcommand{\mrt}{{m}^2_R(T)}
\newcommand{\mk}{{m}^2(k)}
\newcommand{\rhoa}{\rho_1}
\newcommand{\rhob}{\rho_2}
\newcommand{\nua}{\nu_A}
\newcommand{\nub}{\nu_B}
\newcommand{\nuab}{\bar{\nu}_A}
\newcommand{\nubb}{\bar{\nu}_B}
\newcommand{\rhb}{\bar{\rho}}
\newcommand{\rht}{\tilde{\rho}}
\newcommand{\rhta}{\tilde{\rho}_1}
\newcommand{\rhtb}{\tilde{\rho}_2}
\newcommand{\rhz}{\rho_0}
\newcommand{\rhzr}{\rho_{0R}}
\newcommand{\rhza}{\rho_{10}}
\newcommand{\rhzb}{\rho_{20}}
\newcommand{\rhzt}{\tilde{\rho}_0}
\newcommand{\rhzta}{\tilde{\rho}_{10}}
\newcommand{\rhztb}{\tilde{\rho}_{20}}
\newcommand{\rhzk}{\rho_0(k)}
\newcommand{\rhzkt}{\rho_0(k,T)}
\newcommand{\kx}{\kappa}
\newcommand{\kt}{\tilde{\kappa}}
\newcommand{\kk}{\kappa(k)}
\newcommand{\ktk}{\tilde{\kappa}(k)}
\newcommand{\Gammat}{\tilde{\Gamma}}
\newcommand{\Gammak}{\Gamma_k}
\newcommand{\wt}{\tilde{w}}
\newcommand{\be}{\begin{equation}}
\newcommand{\ee}{\end{equation}}
\newcommand{\een}{\end{subequations}}
\newcommand{\ben}{\begin{subequations}}
\newcommand{\beq}{\begin{eqalignno}}
\newcommand{\eeq}{\end{eqalignno}}
\newcommand{\lsim}{\begin{array}{c}<\vspace{-0.32cm}\\\sim\end{array}}
\newcommand{\gsim}{\begin{array}{c}>\vspace{-0.32cm}\\ \sim\end{array}}
\pagestyle{empty}
\noindent
\begin{flushright}
CAU-THP-95-38 \\
CERN-TH/96-67 \\
DESY 96-038 \\
HD-THEP-96-05 \\ 
cond-mat/9603129
\end{flushright} 
\vspace{1cm}
\begin{center}
{ \Large Flow of the Coarse Grained Free Energy \\
for Crossover Phenomena} 
\\ \vspace{1cm}
S. Bornholdt,$^{\rm a,}$\footnote{e-mail: bornholdt@theo-physik.uni-kiel.de}
P. B\"uttner,$^{\rm b,}$\footnote{e-mail: pesebu@x4u2.desy.de}
N. Tetradis$^{\rm c,}$\footnote{e-mail: tetradis@surya11.cern.ch}
and C. Wetterich$^{\rm d,}$\footnote{e-mail: 
c.wetterich@thphys.uni-heidelberg.de}\\
\vspace{1cm}
{}$^{\rm a}$Institut f\"ur Theoretische Physik, Universit\"at Kiel,\\
Leibnizstr. 15, 24098 Kiel, Germany \\
\vspace{0.5cm}
{}$^{\rm b}$II.\ Institut f\"ur Theoretische Physik, Universit\"at Hamburg,\\
Luruper Chaussee 149, 22761 Hamburg, Germany \\
\vspace{0.5cm}
{}$^{\rm c}$CERN, Theory Division, \\ 
CH-1211, Geneva 23, Switzerland \\
\vspace{0.5cm}
{}$^{\rm d}$Institut  f\"ur Theoretische Physik, Universit\"at Heidelberg,\\
Philosophenweg 16, 69120 Heidelberg, Germany\\
\vspace{1cm}
\abstract{
The critical behaviour of a system of two coupled scalar fields in
three dimensions 
is studied within the formalism of the effective average action.
The fixed points of the system are identified and the crossover 
between them is described in detail.
Besides the universal critical behaviour, the flow of the coarse 
grained free energy also describes the approach to scaling.
}
\end{center}
\vspace{1cm}

\clearpage

\setlength{\baselineskip}{15pt}
\setlength{\textwidth}{16cm}
\pagestyle{plain}
\setcounter{page}{1}

\newpage

\setcounter{equation}{0}
\renewcommand{\theequation}{{\bf 1.}\arabic{equation}}

\section*{1. Introduction}

The properties of multicomponent statistical systems 
near the critical temperature of a second-order phase
transition are often characterized by a crossover behaviour. 
Crossover typically occurs if two or more fixed points
are relevant for the description of scaling. Examples
can be found in physical systems 
with small impurities or a small coupling between two 
degrees of freedom. A well developed formalism for the 
description of the universal features of critical 
crossover is based on the renormalization group equations
\cite{wilson}--\cite{reneq}. Nevertheless, the practical
situation in a given experiment may be
quite complicated. The correlation length is large but 
not infinite  in real life and, as a consequence, 
the scaling properties are not perfect. The temperature
dependence of order parameters, susceptibilities, etc., 
often reflects an interplay between the approach to scaling 
and crossover. The first is determined by the flow of couplings
towards a fixed point, whereas the second corresponds to the 
running of couplings from one fixed point to another. 
Standard methods employed for the study of critical phenomena, 
such as for example the $\epsilon$-expansion, are 
often insufficient for the quantitative description 
of such complex situations, which are in addition not completely
governed by universal scaling behaviour. Numerical simulations 
are also difficult in this context, as typically 
several different length scales are involved. 

We propose here the use of a field theoretical description 
in terms of the effective average action $\Gamma_k$. 
This corresponds to a coarse grained free energy where 
fluctuations with momenta larger than $k$ are integrated out.
The flow of $\Gamma_k$ with a change of the 
infrared cutoff scale $k$ is described by an exact non-perturbative
flow equation. This can be solved approximately by appropriate
truncations of the most general form of $\Gamma_k$.
At a given scale $k$ only modes with momenta around 
$k$ contribute to the flow. We specify the microscopic 
physics by the ``initial value'' of $\Gamma_k$ 
at some high momentum scale $\Lambda$. For a given experimental situation
this step involves the rewriting of the problem in terms of 
continuous variables. This part of the problem is not universal 
and involves only the short distance modes. It is not treated
in the present paper. We only mention here that, for lattice models,
it is advantageous to take $\Lambda$ somewhat smaller than
the inverse lattice spacing, so that deviations of the form of 
$\Gamma_{\Lx}$ from translational 
and rotational invariance are not too large. 
Our formalism can be extended to situations without
translational and rotational symmetry. It is, however, 
much simpler for translationally and rotationally 
invariant situations where standard field theory can
be used. In the vicinity of the critical temperature 
$\tcr$ of a second-order phase transition one of the 
couplings parametrizing $\Gamma_{\Lx}$ -- in our case $\dkl$ --
can be used to describe the distance from the critical
hypersurface in the space of couplings,
i.e. $\dkl \sim T - \tcr$. The order parameter, correlation lengths 
etc. can then be studied as functions of $\dkl$.
For crossover situations there exists a second important coupling
-- in our case $x_{\Lx}$ -- which determines to which fixed point
the flow of the couplings is dominantly attracted. The parameter 
$x_{\Lx}$ typically measures the amount of impurities or the size
of a coupling between different components of a system.
Critical quantities, such as the exponent $\nu$ which governs the 
dependence of the correlation length on $\dkl$ 
($\nu = - \partial \ln \xi / \partial \ln|\dkl |$), now depend on the 
ratio $x_{\Lx}/|\dkl |$.

We determine the long distance physics by solving numerically the flow of
$\Gamma_k$, starting at short distances with 
$\Gamma_{\Lx}$ and moving towards $k=0$. For $k=0$ the effective average 
action equals the free energy expressed as a functional of the fields. 
In the language of field theory, $\Gamma_0$ is the generating functional
of the 1PI Green functions, and all physically relevant information can be 
extracted from it.
The method is formulated in arbitrary dimensions and can, in particular,
be applied directly to three-dimensional statistical systems.
Consequently, there is no restriction to the scaling situation
for infinitesimally small $\dkl$. The correlation length and 
similar quantities can be computed for arbitrary $\dkl$ and
$x_{\Lx}$, and the interplay between non-universal and universal
behaviour can be studied by approaching the critical 
temperature ($\dkl \rightarrow 0$).

In this paper we discuss these ideas for a specific model of two
coupled fields with discrete symmetry. The parameter 
$x_{\Lx}$ will be a measure of the strength of the coupling
between the two fields. For the present investigation our main
purpose is not high accuracy for the universal quantities but rather
a demonstration that numerical 
solutions of flow equations can give a realistic picture. We, therefore,
limit ourselves to a relatively rough truncation of $\Gamma_k$.
More involved truncations would give better quantitative 
accuracy but also obscure the presentation 
of the basic mechanisms and complicate the numerical work. 
Such improvements will become relevant once a given
experimental setting with given form of $\Gamma_{\Lx}$ 
is investigated, with emphasis on quantitative accuracy.
They are also needed if one attempts to compute
universal quantities such as crossover exponents with an accuracy comparable 
with other existing methods.

\setcounter{equation}{0}
\renewcommand{\theequation}{{\bf 2.}\arabic{equation}}

\section*{2. The evolution equation for the potential}

We consider a theory of two real scalar fields 
$\chi_a~(a=1,2)$, in
$d$-dimensional Euclidean space, with an 
action $S[\chi]$ invariant under the discrete symmetry
$(1 \leftrightarrow - 1,
2 \leftrightarrow - 2,
1 \leftrightarrow  2)$. 
We specify the action together with some ultraviolet 
momentum cutoff $\Lambda$, so that the theory is properly 
regulated. 
We add to the kinetic term an infrared regulating piece \cite{exact}
\be
\Delta S = \frac{1}{2} \int \frac{d^d q}{(2 \pi)^d}
R_k(q) \chi^*_a(q) \chi^a(q),
\label{twoone} \ee
where $\chi^a(q)$ are the Fourier modes of the scalar fields. 
The function $R_k$  is employed in 
order to prevent the propagation of modes 
with characteristic momenta $q^2 < k^2$. 
This can be achieved, for example,
by the choice 
\be 
R_k(q) = \frac{Z_k q^2 f^2_k(q)}{1 - f^2_k(q)},
~~~~~~~~~
{\rm with}
~~~~ 
f^2_k(q) = \exp \left( - \frac{q^2}{k^2} \right).
\label{twotwo} \ee
The quantity $Z_k$ is an appropriate wave function renormalization
whose precise definition will be given below. 
The modes with $q^2 \gg k^2$ are unaffected by the infrared cutoff,
while the 
low frequency modes with $q^2 \ll k^2$ are cut off,
as $R_k$ acts like a mass term:
$ \lim_{q^2 \rightarrow 0} R_k(q) = Z_k k^2. $
Through the usual Legendre 
transformation we obtain the
generating functional for the 1PI Green functions 
${\tilde \Gamma}_k[\phi^a]$, where $\phi^a$ is the expectation value of the 
field $\chi^a$ in the presence of sources.
The use of the modified propagator for the calculation of 
${\tilde \Gamma}_k$ results in the effective integration of only the 
fluctuations with $q^2 >
k^2$. Finally, the effective average action is 
obtained by subtracting the infrared cutoff piece 
\be
\Gamma_k[\phi^a] = {\tilde \Gamma}_k[\phi^a] -
\frac{1}{2} \int \frac{d^d q}{(2 \pi)^d}
R_k(q) \phi^*_{a}(q) \phi^a(q).
\label{twothree} \ee
For $k$ equal to the ultraviolet cutoff $\Lambda$, $\Gammak$ becomes 
equal 
to the classical action $S$ (no effective integration of modes takes 
place), while for $k \rightarrow 0$ it tends towards the effective action 
$\Gamma$ (all the modes are included),
which is the generating functional of the 1PI Green functions computed 
from $S$ without infrared cutoff.
For intermediate values of $k$ the effective average action 
realizes the concept of a coarse grained free energy \cite{langer}.

The interpolation of $\Gammak$ between the classical and the 
effective action makes it a very useful field theoretical tool.
The means for practical calculations is provided by an exact 
flow equation,
which describes the response of the 
effective average action to variations of the infrared cutoff 
($t=\ln (k/\Lambda)$) \cite{exact}
\be
\frac{\partial}{\partial t} \Gammak[\phi]
= \frac{1}{2} {\rm Tr} \left\lbrace (\Gammak^{(2)}[\phi] + R_k)^{-1} 
\frac{\partial}{\partial t} R_k \right\rbrace. 
 \label{twofour} \ee 
Here $\Gammak^{(2)}$ is the second functional derivative of the effective 
average action with respect to $\phi^a$. 

For the solution of eq. (\ref{twofour}) 
an efficient truncation scheme has to be developed.
The form of the effective average action is constrained by the 
$(1 \leftrightarrow - 1,
2 \leftrightarrow - 2,
1 \leftrightarrow  2)$ 
symmetry. 
However, there is still an infinite number of invariants 
to be considered. 
Throughout this paper we work with  
an approximation which 
keeps only a standard kinetic term in the effective average 
action
\be
\Gammak = 
\int d^dx \left\lbrace U_k(\rhoa, \rhob) 
+ \frac{1}{2} Z_k \partial^{\mu} \phi_a 
\partial_{\mu} \phi^a \right\rbrace,
\label{twofive} \ee
and neglect all invariants which involve more derivatives
of the fields. 
We have used the definition $\rho_a = \frac{1}{2} \phi_a^2$.
The parameter appearing in eq. (\ref{twotwo})
can be identified with the wave function renormalization $Z_k$. 
The justification for our approximation lies in the 
smallness of the anomalous dimension,
which is expected to be $\eta \simeq 0.04$
for the three-dimensional theory. 
We estimate the corrections arising from the 
inclusion of higher derivative terms to 
be of the same order as $\eta$.
In order to obtain an evolution equation for the effective average 
potential $U_k$ 
from eq. (\ref{twofour}), 
we have to expand around a constant background 
field configuration (so that the derivative terms in the parametrization
(\ref{twofive}) do not contribute to the l.h.s. of eq. (\ref{twofour})). 
Equation (\ref{twofour}) then gives \cite{christof2,indices} 
\be
\frac{\partial}{\partial t} U_k(\rhoa,\rhob) = 
\frac{1}{2} \int \frac{d^d q}{(2 \pi)^d} 
\left( \frac{1}{P(q^2) + M^2_1} 
+\frac{1}{P(q^2) + M^2_2} \right)
\frac{\partial}{\partial t} R_k(q).
\label{twosix} \ee
$P(q^2)$ results from
the combination of 
the standard kinetic contribution $Z_k q^2$ 
and the regulating term $R_k$ 
into an effective 
inverse propagator (for massless fields) 
\be
P(q^2) = Z_k q^2 + R_k = \frac{Z_k q^2}{1 - f^2_k(q)},
\label{twoseven} \ee
with $f^2_k(q)$ given in eq. (\ref{twotwo}); 
$M^2_{1,2}$ are the eigenvalues of the mass matrix at the
point $(\rhoa,\rhob)$ \cite{stefan4d,stefan3d}:
\be
M^2_{1,2} = \frac{1}{2} \left\lbrace 
 U_1 + U_2 +2 U_{11} \rhoa + 2 U_{22} \rhob 
\pm  \left[ 
(U_1 - U_2 +2 U_{11} \rhoa - 2 U_{22} \rhob )^2 
+ 16 U^2_{12} \rhoa \rhob \right]^{\frac{1}{2}} \right\rbrace, 
\label{twoeight} \ee
and we have introduced the
notation $U_1 = {\partial U_k}/{\partial \rhoa}$,
$U_{12} = {\partial^2 U_k}/{\partial \rhoa \partial \rhob}$, etc. 

The fixed point structure of the theory is more easily identified if
we use the dimensionless renormalized parameters
\beq
\rht_a= &~Z_k k^{2-d} \rho_a
\nonumber \\
u_k(\rht_1,\rht_2) = &~k^{-d} U_k(\rhoa,\rhob) 
\nonumber \\
\mt^2_a = &~Z_k^{-1} k^{-2} M^2_a. 
\label{twonine} \eeq
The rescaled mass eigenvalues $\mt^2_{1,2}$ are related to
$u_k,\rht_1,\rht_2$ through expressions analogous to eqs.
(\ref{twoeight}), with 
$u_1 = {\partial u_k}/{\partial \rht_1}$, etc.
The evolution equation (\ref{twosix}) can now be written in
the scale-independent form 
\be
\frac{\partial}{\partial t} u(\rhta,\rhtb) =
-d u +(d-2+ \eta) (\rht_1 u_1 + \rht_2 u_2) 
- v_d  L^d_0(\mt^2_1)
- v_d  L^d_0(\mt^2_2),
\label{twoten} \ee
with 
\be 
v_d^{-1} = 2^{d+1} \pi^{\frac{d}{2}} \Gamma \left( \frac{d}{2} \right).
\label{twovd} \ee
The functions $L^d_0(w)$ are given by
\be
L^d_0(w) = -2 \int_0^{\infty} dy y^{\frac{d}{2}+1}
\frac{\exp(-y)}{\left[ 1-\exp(-y) \right]^2}
\left[ \frac{y}{1-\exp(-y)} + w \right]^{-1}.
\label{twoeleven} \ee
We also employ the functions 
\beq 
L^d_1(w) = - &\frac{\partial}{\partial w} L^d_0(w) 
\nonumber \\
 L^d_{n+1}(w) = &- \frac{1}{n} \frac{\partial}{\partial w} L^d_n(w)
~~~~~~~~~~~~~~~~~~{\rm for}~~ n \geq 1
\label{twothirteen} \eeq
in the following. These functions are negative 
for all values of $w$. 
Also $|L^d_n(w)|$ are monotonically decreasing for increasing $w$
and introduce threshold behaviour in the evolution. 
For large values of $\mt^2_a \sim M^2_a /k^2$ 
the last two terms in eq. (\ref{twoten}) vanish and the evolution
of $U_k$ stops. The above functions have been extensively 
discussed in refs. \cite{christof2,indices,convex}. 
As they have no simple analytical form, we use numerical 
fits in the following. 

The evolution of the wave function renormalization $Z_k$, which gives the
anomalous dimension 
\be
\partial (\ln Z_k) / \partial t = - \eta,
\label{twofourteen} \ee  
can be computed from the exact flow equation (\ref{twofour}) by 
expanding around a background field configuration with a small 
momentum dependence \cite{indices}.
We work in three dimensions 
and use the approximate value $\eta=0.04$. 
Within the accuracy of our results, this value 
is consistent with the values of $\eta$ in both the Ising and 
Heisenberg models, which are relevant near the fixed points. 
We take into account threshold effects in the evolution of $Z_k$ 
and the vanishing of $\eta$ in the symmetric regime 
(when the minimum of $U_k$ is at the origin) \cite{indices}
by using $\eta=0.04$ as long as the theory is in the regime 
with spontaneous symmetry breaking and 
the mass eigenvalues are in the interval $0.01 \leq \mt_a^2 \leq 100$.
Otherwise we set $\eta=0$. 
We also point out that we have neglected a term proportional to 
$\eta$ in the r.h.s.  of (\ref{twoten}), which comes from the 
$t$-derivative acting on $Z_k$ inside $R_k$ in the r.h.s. of 
eq. (\ref{twosix}). This is consistent with our approximate 
treatment of the wave function renormalization effects and is 
justified by the smallness of $\eta$.

\setcounter{equation}{0}
\renewcommand{\theequation}{{\bf 3.}\arabic{equation}}

\section*{3. The fixed point structure}

From this point on we concentrate on the  three-dimensional
theory.  
Equation
(\ref{twoten}), with $d=3$, 
is the master equation for our investigation.
It is a non-linear partial differential equation 
for three independent variables.
A numerical solution is possible 
(see refs. \cite{num,eos} for the solution of the 
evolution equation for the $O(N)$-symmetric scalar theory). 
However, the presence of three independent variables complicates
the numerical treatment. 
For this reason we resort to a different approximation scheme.
We first introduce a Taylor expansion of
$U_k(\rhoa,\rhob)$ around a certain point (usually the minimum
of the potential or the origin). 
This turns eq. (\ref{twosix}) into an infinite system of
ordinary (coupled) differential equations for
the $k$-dependence of the point around which the expansion is 
performed 
and the derivatives of the 
potential, with 
independent variable $t=\ln (k/\Lambda)$. 
We solve this system approximately
by truncating at a finite number of derivatives. 
This approach has been used in the past for the study of 
the $O(N)$-symmetric scalar theory. 
It has provided a full, detailed picture of the 
high-temperature phase transition for the four-dimensional theory
\cite{transition,indices,largen,review}, with accurate determination 
(at the few \% level) of 
such non-trivial quantities as the critical 
exponents \cite{indices}. 
An estimate \cite{tim1} of the residual errors for high 
level truncations indicates that
they are smaller than the uncertainties introduced 
by the imprecise treatment of the wave function 
renormalization effects. 
This has been confirmed by comparison with the numerical 
solution of the untruncated evolution equation for the potential 
\cite{num,eos}. 
The same approach has been used in ref. \cite{stefan3d} 
for the study of the
high-temperature phase transition in a four-dimensional 
theory of two scalars with a 
$(1 \leftrightarrow - 1,
2 \leftrightarrow - 2,
1 \leftrightarrow  2)$ symmetry. 
The truncation employed there kept up to second derivatives 
of the potential $U_{11}, U_{12}, U_{22}$.
The present work is a detailed investigation of the crossover phenomena 
in the effective three-dimensional theory   
near the critical temperature. 
We use an extended 
truncation which keeps up to the third derivatives 
of the potential $U_{111}, U_{122}$, etc.
We also note that this theory belongs in the 
same universality class as the statistical two-component spin 
systems with 
cubic anisotropy, whose universal behaviour has been studied
in \cite{aharony}--\cite{amit}.

Let us first parametrize the potential
by its derivatives at the origin (S regime)
\beq
\mb^2(k) &= U_1(0) = U_2(0) \nonumber \\  
\lb(k) &= U_{11}(0) = U_{22}(0)~~~~~~~~~~  
\gb(k) = U_{12}(0) \nonumber \\
\nuab(k) &= U_{111}(0) = U_{222}(0)~~~~~~~~  
\nubb(k) = U_{112}(0) = U_{122}(0), 
\label{threeone} \eeq
where the equality of derivatives 
is imposed  by the 
discete symmetry of the theory. 
Our approximation amounts to using, for the potential, the
ansatz
\be
U_k(\rhoa,\rhob) = \mb^2(\rhoa+\rhob)
+\frac{\lb}{2} \left( \rhoa^2 + \rhob^2 \right) 
+\gb \rhoa \rhob
+\frac{\nuab}{6} \left( \rhoa^3 + \rhob^3 \right) 
+\frac{\nubb}{2} \left( \rhoa^2 \rhob + \rhoa \rhob^2 \right). 
\label{threetwo} \ee
Evolution equations for the above parameters are obtained 
by taking derivatives of
eq.~(\ref{twosix}) with respect
to $\rho_{1,2}$.
It is convenient to define the dimensionless couplings
\beq
\mt^2(k) = &Z_k^{-1}~k^{-2}~\mb^2(k) = k^{-2}~m^2(k)
\nonumber \\
\lt(k) = &Z_k^{-2}~k^{-1}~\lb(k) = k^{-1}~\lx(k)   
~~~~~~~
\gt(k) = Z_k^{-2}~k^{-1}~\gb(k) = k^{-1}~g(k)
\nonumber \\
\nua(k) = &Z_k^{-3}~\nuab(k) 
~~~~~~~~~~~~~~~~~~~~~~~~~
\nub(k) = Z_k^{-3}~\nubb(k)
\label{threethree} \eeq
The dimensionless potential $u_k(\rhta,\rhtb)$
defined in eqs. (\ref{twonine}) 
can be expressed in complete analogy to eq. (\ref{threetwo})
in terms of 
these dimensionless quantities. 
The evolution equations for the above quantites have a
scale-invariant form, in the sense that the r.h.s. does not
explicitly involve a dependence on $k$ 
\beq
\frac{d \mt^2}{d t} = 
&(-2+\eta) \mt^2 
+ v_3   (3 \lt+ \gt)  L_1^3(\mt^2)
\label{threefour} \\
\frac{d \lt}{d t} = 
&(-1 + 2 \eta) \lt 
- v_3  (9 \lt^2 + \gt^2)  L_2^3(\mt^2)
+ v_3  (5 \nua + \nub)  L_1^3(\mt^2)
\label{threefive} \\
\frac{d \gt}{d t} = 
&(-1 + 2 \eta) \gt 
- v_3  (6 \lt \gt + 4 \gt^2)  L_2^3(\mt^2)
+ 6 v_3  \nub  L_1^3(\mt^2)
\label{threesix} \\
\frac{d \nua}{d t} = 
&3 \eta \nua 
+ 2 v_3  (27 \lt^3 + \gt^3)  L_3^3(\mt^2)
- 3 v_3  (15 \lt \nua + \gt \nub)   L_2^3(\mt^2)
\label{threeseven} \\
\frac{d \nub}{d t} = 
&3 \eta \nub
+ 2 v_3  (9 \lt^2 \gt + 15 \lt \gt^2 + 4 \gt^3)  L_3^3(\mt^2)
- v_3  (21 \lt \nub + 5 \gt \nua + 22 \gt \nub)   L_2^3(\mt^2),
\nonumber \\
~&~ 
\label{threeeight} 
\eeq
where $v_3 = 1/8\pi^2$.
They automatically preserve the discrete symmetry.

We are interested in the 
fixed points of the last set of 
equations.
First there is the ultraviolet attractive Gaussian fixed point (G) 
with $\mt^2,\lt,\gt,\nua,\nub=0$.
There are also three fixed points with at least one
infrared attractive direction \cite{aharony}--\cite{amit}.
They all appear for $\mt^2 < 0$, $\lt,\gt,\nua,\nub \geq 0$, which 
means that the minimum of $u_k(\rhta,\rhtb)$ is away from the
origin. 
(The exact values are 
not important since the discussion in this section concentrates
on the qualitative behaviour only.)
For their identification we use their standard names 
in statistical physics \cite{aharony}--\cite{amit}:    \\
a) The Heisenberg fixed point (H) has 
$\gt=\lt$, $\nub=\nua$ and corresponds to 
a theory with symmetry increased to $O(2)$. The potential $u_k(\rhta,\rhtb)$
has
a series of degenerate minima at $\rhzta+\rhztb= \rhzt$. \\
b) The Ising fixed point (I) has $\gt=\nub=0$ and corresponds to two
disconnected $Z_2$-symmetric theories. 
The potential $u_k(\rhta,\rhtb)$ has
minima at $\rhzta=\rhztb= \rhzt/2$
(for positive and 
negative values of $\phi_{1,2}$). \\
c) The Cubic fixed point (C) has $\gt=3 \lt$, $\nub=5 \nua$ 
and corresponds to two disconnected $Z_2$-symmetric theories, 
if the fields are redefined 
according to 
\be
\phi'_1 = \frac{1}{\sqrt{2}} (\phi_1 + \phi_2),
~~~~~~~~~~~~~
\phi'_2 = \frac{1}{\sqrt{2}} (\phi_1 - \phi_2).
\label{threenine} \ee
The potential $u_k(\rhta,\rhtb)$ has
minima at $\rhzta= \rhzt$, $\rhztb=0$ and $\rhztb= \rhzt$, $\rhzta=0$. \\
All these points are infrared unstable in the $\mt^2$ direction 
and are located on a critical surface 
$\mt^2_{cr} = \mt^2_{cr}(\lt,\gt,\nua,\nub) < 0$. Solutions of the 
evolution equations, which start above the critical surface,
with $\mt^2 > \mt^2_{cr}$, flow towards the region of positive 
$\mt^2$ for $t \rightarrow -\infty$,
and correspond to theories in the symmetric phase.
Solutions with
$\mt^2 < \mt^2_{cr}$ flow deep into the region of negative $\mt^2$ 
and correspond to theories in the phase 
with spontaneous symmetry breaking.

The relative stability of the fixed points on the critical 
surface determines which one governs the dynamics of the 
phase transition in the immediate vicinity of the critical 
temperature (or mass).
It has been discussed in ref. \cite{stefan3d}, in a 
truncation that neglects derivatives of the potential
higher than the second (and corresponds to $\nua,\nub=0$). 
All three
fixed points are attractive in the $\lt$ direction. However, the
Ising and Cubic fixed points are repulsive in the $\gt$ direction,
while the Heisenberg fixed point is totally attractive.
There are four disconnected regions: \\
a) $3 \lt > \gt> \lt$ : The trajectories flow away from the Cubic 
towards the Heisenberg fixed point. \\
b) $\lt> \gt > 0$ : The trajectories flow away from the Ising
towards the Heisenberg fixed point. \\ 
For parameters in the above two regions
we expect second-order phase transitions,
with critical dynamics governed by the three fixed points.
\\
c) $\gt > 3 \lt$ : The trajectories flow 
away from the Cubic fixed point into a region of large $\gt$ and
small $\lt$. Eventually $\lt$ turns negative at a finite value
of $k$. 
At this point an instability arises which signals the presence of 
a first-order phase transition.\\
d) $\gt < 0$ : The trajectories flow 
away from the Ising fixed point and 
cross the line $\gt = -\lt $ at a finite $k$.
This again implies the presence of an instability whose 
true nature is related to a first-order phase transition. \\
Flows that start on the lines $\gt=0, \lt, 3 \lt$ remain on these lines.
No trajectories exist which connect the four regions  
$\gt > 3 \lt$, $3 \lt > \gt> \lt$, $\lt> \gt > 0$, $\gt < 0$.

We are interested in studying in detail the crossover 
from the Ising to the Heisenberg fixed point in the region
$\lt> \gt > 0$. The region $3 \lt > \gt> \lt$ 
can be mapped on the region $\lt> \gt > 0$
through a redefinition of the fields \cite{stefan3d}
according to eqs. (\ref{threenine}). This shows that
the Ising and Cubic fixed points lead to 
an identical universal behaviour (and 
therefore to identical universal quantities, such as critical
exponents and crossover curves).

\setcounter{equation}{0}
\renewcommand{\theequation}{{\bf 4.}\arabic{equation}}

\section*{4. The crossover}

We would like to study the crossover from the 
Ising to the Heisenberg fixed point. In the vicinity of these
fixed points the potential 
$u_k(\rhta,\rhtb)$ has
minima at $\rhzta=\rhztb= \rhzt/2 >0$ (for positive and 
negative values of $\phi_{1,2}$). In order to improve our
quantitative accuracy in this regime 
(which we call the M regime), it is convenient to parametrize 
the potential in terms of its derivatives at the minimum. 
For this reason we define the parameters 
\beq
\lb(k) &= U_{11}(\rhz) = U_{22}(\rhz)~~~~~~~~~~  
\gb(k) = U_{12}(\rhz) \nonumber \\
\nuab(k) &= U_{111}(\rhz) = U_{222}(\rhz)~~~~~~~~  
\nubb(k) = U_{112}(\rhz) = U_{122}(\rhz), 
\label{fourone} \eeq
where $\rhza=\rhzb= \rhz/2$ is the minimum of $U_k(\rhoa,\rhob)$, 
related to the minimum of $u_k(\rhta,\rhtb)$
through the first of eqs. (\ref{twonine}). 
The equality of derivatives 
is imposed  by the 
$(1 \leftrightarrow - 1,
2 \leftrightarrow - 2,
1 \leftrightarrow  2)$ 
symmetry of the theory. 
The first derivatives $U_{1}(\rhz) = U_{2}(\rhz)$ are zero for
non-zero $\rhz$.
Our approximation amounts to using for the potential the
ansatz
\beq
U_k(\rhoa,\rhob) = 
&~\frac{\lb}{2} \left[ 
\left( \rhoa - \frac{\rhz}{2} \right)^2 
+\left( \rhob - \frac{\rhz}{2} \right)^2 
\right]
+\gb 
\left( \rhoa - \frac{\rhz}{2} \right)
\left( \rhob - \frac{\rhz}{2} \right)
\nonumber \\
&+\frac{\nuab}{6} 
\left[ 
\left( \rhoa - \frac{\rhz}{2} \right)^3 
+\left( \rhob - \frac{\rhz}{2} \right)^3 
\right]
\nonumber \\
&+\frac{\nubb}{2} 
\left[ 
\left( \rhoa - \frac{\rhz}{2} \right)^2 
\left( \rhob - \frac{\rhz}{2} \right) 
+\left( \rhoa - \frac{\rhz}{2} \right) 
\left( \rhob - \frac{\rhz}{2} \right)^2 
\right].
\label{fourtwo} \eeq
We define dimensionless couplings according to eqs. (\ref{threethree})
for the derivatives and 
\be
\kx(k) = Z_k k^{-1} \rhz(k)
\label{fourthree} \ee
for the minimum.
The evolution equations for these couplings are lengthy 
and can be found in the appendix.
They automatically preserve the 
discrete symmetry. The various fixed points are now
located on a critical surface 
$\kx^2_{cr} = \kx^2_{cr}(\lt,\gt,\nua,\nub)$, which is unstable in the 
$\kx$ direction and separates the phase with spontaneous symmetry
breaking from the symmetric one. 
Solutions of the evolution 
equation starting with $\kx > \kx_{cr}$ flow towards the region of 
large $\kx$ and correspond to the phase with symmetry breaking. 
Eventually the minimum of the potential settles down at a finite value
\be
\rhzr = \lim_{k \rightarrow 0} Z_k \rhz(k) = \lim_{k \rightarrow 0} 
k \kx(k). 
\label{fourfour} \ee
Solutions
starting with $\kx < \kx_{cr}$ lead to 
$\kx(k) = 0$ at some finite $k_s$. 
From this point on, the system 
is in the symmetric regime 
(S regime) and we use a parametrization of the potential 
according to eqs. (\ref{threeone}) and (\ref{threetwo}), the
rescaled variables defined in eqs. (\ref{threethree}), and 
the evolution   
equations (\ref{threefour})--(\ref{threeeight}).
Finally, if the evolution starts on the critical surface 
the flows never lead out of it. Instead they eventually approach 
the most attractive fixed point on the critical surface. 
This means that for
$k \rightarrow 0$ the renormalized minimum of the potential
is given by eq. (\ref{fourfour}) with finite $\kx$.
Therefore, $\rhzr=0$ and the critical surface corresponds to a 
second-order phase transition.

We numerically integrate the evolution equations by
starting at some high momentum scale (ultraviolet cutoff)
$k=\Lx$ ($t=0$) where the potential $U_{\Lx}$ is equal to the 
classical potential $V$. 
We consider $U_{\Lx}$ quartic in the fields, and therefore given
by eq. (\ref{fourtwo}) with 
$\nuab(\Lx)=\nubb(\Lx)=0$.
For the wave function renormalization we start with $Z_{\Lx}=1$. 
We fine-tune $\rhz(\Lx)/\Lx=\kx(\Lx)$, so that 
$|\dkl|=|\kx(\Lx)-\kx_{cr}| \ll 1$, and the system is 
very close to the critical surface.
The subsequent 
evolution moves the parameters along 
the critical surface and close to various fixed points 
depending on their relative stability. 
We define the quantity 
\be
x(k) = \frac{\gb(k)}{\lb(k)} -1=\frac{\gt(k)}{\lt(k)}-1 
\label{fourfive} \ee
in order to parametrize the distance from the Heisenberg fixed point.
We have $x_I=-1$ at the Ising fixed point and $x_H=0$ at the Heisenberg one.
We can arrange for the evolution to approach first the 
Ising fixed point by 
selecting $\dxl=x(\Lx)+1 \ll 1$.
Then $x(k)$ remains very close to its initial value while the 
running couplings approach their values at 
the Ising fixed point
\be
\kx_I=8.57 \times 10^{-2},~~~~~
\lt_I=8.05,~~~~~
\gt_I=0,~~~~~
(\nua)_I=75.8,~~~~~
(\nub)_I=0.   
\label{foursix} \ee
Near the fixed point the evolution slows down, as the $\beta$-functions
are almost zero. However, the Ising fixed point has two unstable directions
and the system will eventually move away from it. 
One of these directions leads away from the critical surface, while the other 
is directed towards the Heisenberg fixed point. 
By selecting the starting point of the evolution so that 
$\dxl \gg |\dkl|$, we can arrange for the flow to stay close to 
the critical surface while moving from 
one fixed point to the other. Eventually 
the running couplings approach their values at 
the Heisenberg fixed point
\be
\kx_H=6.02 \times 10^{-2},~~~~~
\lt_H=7.32,~~~~~
\gt_H=7.32,~~~~~
(\nua)_H=56.4,~~~~~
(\nub)_H=56.4.   
\label{fourseven} \ee
The trajectory in coupling space connecting the two fixed points 
is depicted in fig. 1, while the $\beta$-functions 
for the various couplings
along the trajectory 
are plotted in fig. 2. It is clear that the evolution is very slow near
the fixed points where the $\beta$-functions are zero and fast in the
middle of the trajectory. 
As long as $|\dkl| / \dxl \ll 1$ the system stays close to the critical
surface for a sufficiently long ``time'' $t$ for the Heisenberg fixed 
point to be reached. However, if the ratio $|\dkl| / \dxl$
is increased the system leaves the 
critical surface at some earlier point of the evolution along the 
critical trajectory of fig. 1, and moves towards the phase with
symmetry breaking or towards 
the symmetric one. For $\dkl < 0$ it runs into
the S regime at some finite $k_s$ and eventually settles down in the 
symmetric phase with renormalized parameters \cite{indices,stefan3d}
\beq
m^2_R = &\lim_{k \rightarrow 0} m^2(k) = \lim_{k \rightarrow 0} k^2 \mt^2(k) 
\nonumber \\
\lx_R = &\lim_{k \rightarrow 0} \lx(k) = \lim_{k \rightarrow 0} k \lt(k) 
~~~~~~~
g_R = \lim_{k \rightarrow 0} g(k) = \lim_{k \rightarrow 0} k \gt(k) 
~~~~~~~
x_R = \frac{g_R}{\lx_R}-1
\nonumber \\
\nu_{AR} = &\lim_{k \rightarrow 0} \nua(k)
~~~~~~~~~~~~~~~~~~~~~ 
\nu_{BR} = \lim_{k \rightarrow 0} \nub(k). 
\label{foureight} \eeq
For $\dkl > 0$ the system settles down in the phase with 
symmetry breaking, with the minimum of the 
potential given by eq. (\ref{fourfour}) and the various
couplings at the minimum defined in analogy with eqs.
(\ref{foureight})
\footnote{In the vicinity of the Heisenberg fixed point 
the presence of a massless Goldstone mode results in zero values 
for $\lx_R$, $g_R$ in the limit $k \rightarrow 0$. 
Quartic couplings 
can be defined 
at non-zero values of $k$ \cite{transition,indices}.}.

The singular behaviour of the theory in the critical region near
the phase transition can be parametrized in terms of critical
exponents. For example, the renormalized mass in the symmetric
phase can be expressed as 
\be
m^2_R \propto |\dkl|^{2\nu}.
\label{fournine} \ee
The parameter $\nu$ can be computed as the slope of the 
function $\ln m^2_R = f \left( \ln|\dkl| \right)$.
If the system is in the vicinity of a fixed point,  
$\nu$ is constant and can be identified with
the corresponding critical exponent. Moreover, this exponent
can be related \cite{wilson,wegner,goldenfeld}
to the most negative eigenvalue $\lx_1$ of the 
matrix $\beta_{i,j}=\partial \beta_i/ \partial a_j$ (with
$a_j=\kx,\lt,\gt,\nua,\nub$), which describes 
the growth 
of small perturbations 
$\dk = \kx-\kx_{fp}$,
$\dlt = \lt-\lt_{fp}$, etc.,
around the fixed point, i.e. 
\be
\nu = - \frac{1}{\lx_1}.
\label{fourten} \ee
In fig. 3 we plot the three smallest eigenvalues of the matrix 
$\beta_{i,j}$ for small perturbations 
$\dk = \kx-\kx_{tr}$,
$\dlt = \lt-\lt_{tr}$, etc.,
around the trajectory of fig. 1. 
For this calculation we employ the 
$\beta$-functions given in the appendix.
We observe that the smallest eigenvalue remains negative
along the whole trajectory. It corresponds to the 
unstable direction away from the critical surface.
From its values at $x=-1$ and 
$x=0$, making use of eq.~(\ref{fourten}),
we can infer the critical exponent $\nu$ of the
Ising and Heisenberg models
\beq
\nu_I = &0.700
\label{foureleven} \\
\nu_H = &0.733.
\label{fourtwelve} \eeq
The above values deviate at the 10\% level from the 
more accurate results obtained through the $\epsilon$-expansion
or summed perturbation series 
($\nu_I=0.631(2), \nu_H=0.670(3)$ \cite{zinn}), 
or from results obtained with the same method and
more extended truncations \cite{indices}. 
This is not surprising, as
our very rough treatment of the wave function renormalization generates
uncertainties of the order of the anomalous dimension $\eta$, as we
mentioned at the end of section 2. 
We observe that the difference $\nu_H-\nu_I=0.033$ compares rather well
with the high precision estimate 0.039.  
We have verified the above values for the critical exponents 
by the explicit computation of the function 
$\ln m^2_R = f \left( \ln|\dkl| \right)$. For this we integrated the 
evolution equations for theories with  
$|\dkl|, \dxl \ll 1$, as we discussed in the previous paragraphs. 
For $|\dkl| \gg \dxl$ the evolution approaches the Ising fixed 
point and slows down for a long ``time'' $t$. As a result the 
system loses memory of the initial conditions of the running 
(the classical parameters of the theory). Eventually the system
leaves the critical surface in the vicinity of the Ising fixed point.
The function 
$\ln m^2_R = f \left( \ln|\dkl| \right)$ has a constant slope, 
which reproduces the value of eq. (\ref{foureleven})
for the exponent $\nu$. For theories
with a smaller $|\dkl|/\dxl$ ratio the evolution first approaches 
the Ising fixed point and then deviates from it along the critical 
surface. Finally it leaves the critical surface at some point 
along the trajectory of fig. 1 connecting the two fixed points. 
The function 
$\ln m^2_R = f \left( \ln|\dkl| \right)$ now has a variable slope, since
the growth of deviations from the critical surface depends on the average
of the eigenvalue $\lx_1$ along the part of the trajectory covered by 
the evolution. Only for  
 $|\dkl| \ll \dxl$ does the slope become constant again. For these theories
the flow first moves from the Ising to the Heisenberg fixed point,  
and then spends a long ``time'' $t$ in its vicinity. During this
``time'' it loses memory of the previous evolution, before
eventually leaving the critical surface.
The calculation of the exponent $\nu$ reproduces the value of
eq. (\ref{fourtwelve}). The error in the above values of $\nu$ 
originating in uncertainties of the numerical integration 
is less that $1\%$ and, therefore, much smaller than the 
$10\%$ error resulting from the truncations of the evolution equation.

We now turn to the second smallest eigenvalue of fig. 3, which
changes sign along the trajectory connecting the two fixed points. 
Its negative sign at $x=-1$ indicates that the Ising fixed point
has two unstable directions, while its positive value at $x=0$
indicates that the Heisenberg fixed point has only one.
The eigenvalue $\lx_2$ corresponds to the
direction along the critical surface, which leads from the Ising to the 
more stable Heisenberg fixed point. 
Its value at $x=-1$ determines the growth of perturbations that
couple the two disconnected $Z_2$-symmetric theories
of the Ising fixed point. 
For the renormalized theory 
these perturbations are parametrized in terms of 
the dimensionless quantity $x_R+1=g_R/\lx_R$. A
crossover exponent $\varphi$ can be defined through the
relation
\be
x_R+1 \propto |\dkl|^{-\varphi}.
\label{fourthirteen} \ee
In the vicinity of the Ising fixed point this exponent is related to 
the two smallest eigenvalues according to \cite{wilson,wegner,goldenfeld}
\be
\varphi = \frac{\lx_2}{\lx_1} = 0.032. 
\label{fourfourteen} \ee
Alternatively $\varphi$ can be computed from the slope of the
function 
$\ln (x_R+1) = f \left( \ln|\dkl| \right)$. The two results are in 
agreement. 
Comparison with more accurate results obtained through the 
$\epsilon$-expansion ($\varphi=0.110(5)$, \cite{aharony,amit}) 
indicates that our approximations result in limited accuracy for such a 
small quantity. In fact, the eigenvalue $|\lx_2|$ is accidentally
very small on the scale of the other eigenvalues $|\lx_1|$ and $|\lx_3|$ 
(cf. fig. 3). A modest truncation error can easily change it by
a factor 3 or 4. It is not very difficult to increase the 
truncation level
in order to reach a few per cent accuracy for $\nu$ \cite{indices}
and correspondingly for $\varphi$.
(We guess that a truncation at the level of ref. \cite{indices}
should yield eigenvalues $\lx_i$ with an accuracy of 
$\simeq 0.05$ and, therefore, an uncertainty in the $\varphi$
value $\simeq 0.03$.)
A precise estimate of $\varphi$
is, however, not the purpose of this paper. As mentioned in the 
introduction we rather want to demonstrate
that our method is capable of covering the 
whole crossover region between the two fixed points.

In fig. 4 we plot the difference of the
``effective'' exponent $\nu= \partial 
\ln m_R / \partial \ln |\dkl|$ from its value at the Ising fixed point
as a function of $x(\Lx)$. 
For this we integrate numerically the evolution equations by
starting at the ultraviolet cutoff
$k=\Lx$ ($t=0$) where the potential $U_{\Lx}$ is equal to the 
classical potential $V$. 
We consider $U_{\Lx}$ quartic in the fields, with 
$\bar{\lx}(\Lx)=5$ and fine-tune
$\kx(\Lx)$ so that the flows approach the critical surface.
By varying $|\dkl|$ we obtain theories 
with various values of the renormalized mass.
In fig. 4 we also plot the renormalized 
parameter $x_R$ defined in eqs. (\ref{foureight}).
The curves (a) correspond to theories
with renormalized mass 
$m_R/\Lx \sim 10^{-4}$ -- $10^{-3}$, while the curves (b) 
correspond to theories with renormalized mass
$m_R/\Lx \sim 10^{-5}$ -- $10^{-4}$. The presence of small kinks in
these curves gives a feeling of the typical errors resulting
from the numerical integration -- a less than $1\%$ effect.
It is clear that the evolution results in values for $x$ 
in the limit $k \rightarrow 0$ which are closer to the Heisenberg 
fixed point than the ``initial value'' $x(\Lx)$.
Both curves for $x_R$ are above the dotted line, which corresponds to 
$x_R=x(\Lx)$. 
Theories with smaller renormalized mass stay  
close to the critical surface for longer ``time'' $t$, 
so that the attraction 
of the Heisenberg fixed point is more pronounced. For given
$x(\Lx)$ this results in  
$x_R$ being  closer to zero for curve (b) than for curve (a). Similarly
the value of $\nu$ for curve (b)
is closer to the Heisenberg fixed point value than for curve (a).
We have also included in fig. 4 the line corresponding to
$-\lx^{-1}_1(x)$ (dashed-dotted line), 
with $\lx_1$ the most negative eigenvalue of
the stability matrix plotted in fig. 3. 
For sufficiently small $m_R/\Lx$ the flows quickly approach the critical 
trajectory of fig. 1 at some point $x \simeq x(\Lx)$. 
Then the running follows the critical trajectory 
towards the Heisenberg fixed point, until eventually it 
moves away from the critical surface towards the symmetric regime.
The resulting exponent $\nu$ corresponds to an average 
over the values of $-\lx^{-1}_1(x)$ for the part of the flow 
close to the critical trajectory, along with some non-universal
contributions from the initial and final running. For sufficiently 
small $m_R/\Lx$ (such as for curve (b))
the latter contributions become insignificant.
The resulting critical exponent $\nu$ is larger than the 
value of $-\lx^{-1}_1$ at the point $x(\Lx)$, because of
the running towards the Heisenberg fixed point. For this reason 
the curve (b) for $\nu$ lies above the dash-dotted line.
For the curve (a)
the non-universal initial and final running
give significant contributions, which contaminate the 
universal part along the critical trajectory. 
The resulting ``effective'' exponent $\nu$ has a complicated dependence 
on $x(\Lx)$. 
It should be pointed out that in most experimental situations 
the observed values of 
$m_R/\Lx$ are not sufficiently small for the non-universal
physics to be insignificant. 
It seems difficult to perform investigations 
of crossover based only on the value of the critical exponents.
This would require to disentangle the crossover from
the non-universal behaviour.
Figure 4 is an explicit demonstration
that our method can effectively deal with such situations. 
Starting from a given set of values for the parameters of the theory 
at the short distance scale $\Lx$, both the universal and 
non-universal long distance physics 
can be determined through the integration of the evolution
equations.

\setcounter{equation}{0}
\renewcommand{\theequation}{{\bf 5.}\arabic{equation}}

\section*{5. Conclusions}

We have demonstrated that the flow of the effective average action
$\Gamma_k$ can be used for a description of crossover phenomena over the
whole crossover region. We have computed both universal quantities --
the eigenvalues $\lx_i(x)$ of the stability matrix of small fluctuations
along the crossover curve parametrized by $x$ -- as well as actual
``effective'' exponents (see fig. 4) for given values of
$m_R/\Lx$. Even though an exact non-perturbative flow equation for 
$\Gamma_k$ is available, practical calculations are feasible only for 
truncations of $\Gamma_k$. These approximations have not been
pushed to a high level of accuracy in the present paper. High precision 
calculations within the $\epsilon$-expansion have determined 
the crossover exponent $\varphi$ with very satisfactory accuracy in 
the past. However, for many practical applications the actually observed
critical behaviour in crossover situations is
not only determined by the universal critical behaviour. It is often
a mixture of the flow towards the fixed points and the subsequent 
crossover flow from one fixed point to another. In these 
situations of ``imperfect scaling'', many standard methods, as for example
the $\epsilon$-expansion, cannot be applied easily. 
The flow of the effective average action  
is directly formulated in three dimensions and includes the relevant
degrees of freedom at every scale $k$ (not only for $k$ in the scaling
region). It can be used both in the symmetric phase and in the 
phase with spontaneous symmetry breaking, where the existence of
massless Goldstone bosons hinders the application of many standard 
methods. 
For a given initial form of the ``microscopic action''
$\Gamma_{\Lx}$, we can solve these equations numerically and compute
the quantities of interest. We hope that the present method
will develop in the future into a useful practical tool for quantitative
investigations of crossover problems.

\newpage

\setcounter{equation}{0}
\renewcommand{\theequation}{{\bf A.}\arabic{equation}}

\section*{Appendix: The evolution equations in the M regime}

It is easier to derive the evolution equations for 
the dimensionless derivatives
in eqs. (\ref{threethree}) and the rescaled minimum
of the potential of eqs. (\ref{fourthree}) by making use of 
the evolution equation (\ref{twoten}) for the rescaled 
potential $u_k(\rhta,\rhtb)$ in three dimensions
\beq
\frac{\partial}{\partial t} u(\rhta,\rhtb) = 
&-3 u +(1 + \eta) (\rht_1 u_1 + \rht_2 u_2) 
+ \zeta
\nonumber \\ 
\zeta = &- v_3  L^3_0(\mt^2_1)
- v_3  L^3_0(\mt^2_2),
\label{aone} \eeq
where the rescaled mass eigenvalues are given by 
\beq
\mt^2_{1,2}(\rhta,\rhtb) = 
&\frac{1}{2} \left(
u_1 + u_2 +2 u_{11} \rhta + 2 u_{22} \rhtb
\pm A \right)
\nonumber \\
A = &\left[ 
(u_1 - u_2 +2 u_{11} \rhta - 2 u_{22} \rhtb )^2 
+ 16 u^2_{12} \rhta \rhtb \right]^{\frac{1}{2}}.
\label{atwo} \eeq
The evolution of $\kx(k)$ is obtained by taking the 
total $t$-derivative of the minimization conditions
$\partial u_k / \partial \rhta |_\kx =
\partial u_k / \partial \rhtb |_\kx = 0.$
We obtain
\be
\frac{d \kx}{dt} = - (1 +\eta) \kx - \frac{2}{\lt + \gt} 
\left. \frac{\partial \zeta}{\partial \rhta} \right|_\kx.
\label{athree} \ee 
For the derivatives we find 
\beq
\frac{d \lt}{dt} = &(- 1 +2 \eta) \lt 
+ \left. \frac{\partial^2 \zeta}{\partial \rhta^2} \right|_\kx
- \frac{\nua + \nub}{\lt + \gt} 
\left. \frac{\partial \zeta}{\partial \rhta} \right|_\kx
\nonumber \\
\frac{d \gt}{dt} = &(- 1 +2 \eta) \gt 
+ \left. \frac{\partial^2 \zeta}{\partial \rhta \partial \rhtb} \right|_\kx
- \frac{2 \nub}{\lt + \gt} 
\left. \frac{\partial \zeta}{\partial \rhta} \right|_\kx
\nonumber \\
\frac{d \nua}{dt} = &3 \eta \nua 
+ \left. \frac{\partial^3 \zeta}{\partial \rhta^3} \right|_\kx
\nonumber \\
\frac{d \nub}{dt} = &3 \eta \nub
+ \left. \frac{\partial^3 \zeta}{\partial \rhta^2 \partial \rhtb} \right|_\kx.
\label{afour} \eeq
The necessary $\rht$-derivatives of $\zeta$ are given by
\beq 
\frac{\partial \zeta}{\partial \rhta} =
&~v_3 \left[
\frac{\partial \mt^2_1}{\partial \rhta} L^3_1(\mt^2_1)
+ \frac{\partial \mt^2_2}{\partial \rhta} L^3_1(\mt^2_2)
\right]
\nonumber \\
\frac{\partial^2 \zeta}{\partial \rhta^2} =
&-v_3 \left\lbrace
\left[ \frac{\partial \mt^2_1}{\partial \rhta} \right]^2 
L^3_2(\mt^2_1)
+\left[ \frac{\partial \mt^2_2}{\partial \rhta} \right]^2 
L^3_2(\mt^2_2)
\right\rbrace
\nonumber \\
&+v_3 \left\lbrace
\frac{\partial^2 \mt^2_1}{\partial \rhta^2} 
L^3_1(\mt^2_1)
+\frac{\partial^2 \mt^2_2}{\partial \rhta^2}
L^3_1(\mt^2_2)
\right\rbrace
\nonumber \\
\frac{\partial^2 \zeta}{\partial \rhta \partial \rhtb} =
&-v_3 \left\lbrace
\frac{\partial \mt^2_1}{\partial \rhta} 
\frac{\partial \mt^2_1}{\partial \rhtb} 
L^3_2(\mt^2_1)
+ \frac{\partial \mt^2_2}{\partial \rhta} 
\frac{\partial \mt^2_2}{\partial \rhtb} 
L^3_2(\mt^2_2)
\right\rbrace
\nonumber \\
&+v_3 \left\lbrace
\frac{\partial^2 \mt^2_1}{\partial \rhta \partial \rhtb} 
L^3_1(\mt^2_1)
+\frac{\partial^2 \mt^2_2}{\partial \rhta \partial \rhtb}
L^3_1(\mt^2_2)
\right\rbrace
\nonumber \\
\frac{\partial^3 \zeta}{\partial \rhta^3} =
&~2 v_3 \left\lbrace
\left[ \frac{\partial \mt^2_1}{\partial \rhta} \right]^3 
L^3_3(\mt^2_1)
+ \left[ \frac{\partial \mt^2_2}{\partial \rhta} \right]^3 
L^3_3(\mt^2_2)
\right\rbrace
\nonumber \\
&-3 v_3 \left\lbrace
\frac{\partial \mt^2_1}{\partial \rhta}
\frac{\partial^2 \mt^2_1}{\partial \rhta^2}
L^3_2(\mt^2_1)
+\frac{\partial \mt^2_2}{\partial \rhta}
\frac{\partial^2 \mt^2_2}{\partial \rhta^2} 
L^3_2(\mt^2_2)
\right\rbrace
\nonumber \\
&+v_3 \left\lbrace
\frac{\partial^3 \mt^2_1}{\partial \rhta^3}
L^3_1(\mt^2_1)
+\frac{\partial^3 \mt^2_2}{\partial \rhta^3}
L^3_1(\mt^2_2)
\right\rbrace
\nonumber \\
\frac{\partial^3 \zeta}{\partial \rhta^2 \partial \rhtb} =
&~2 v_3 \left\lbrace
\left[ \frac{\partial \mt^2_1}{\partial \rhta} \right]^2 
\frac{\partial \mt^2_1}{\partial \rhtb}
L^3_3(\mt^2_1)
+\left[ \frac{\partial \mt^2_2}{\partial \rhta} \right]^2 
\frac{\partial \mt^2_2}{\partial \rhtb}
L^3_3(\mt^2_2)
\right\rbrace
\nonumber \\
&-2 v_3 \left\lbrace
\frac{\partial \mt^2_1}{\partial \rhta} 
\frac{\partial^2 \mt^2_1}{\partial \rhta \partial \rhtb} 
L^3_2(\mt^2_1)
+ \frac{\partial \mt^2_2}{\partial \rhta} 
\frac{\partial^2 \mt^2_2}{\partial \rhta \partial \rhtb} 
L^3_2(\mt^2_2)
\right\rbrace
\nonumber \\
&- v_3 \left\lbrace
\frac{\partial^2 \mt^2_1}{\partial \rhta^2} 
\frac{\partial \mt^2_1}{\partial \rhtb} 
L^3_2(\mt^2_1)
+ \frac{\partial^2 \mt^2_2}{\partial \rhta^2} 
\frac{\partial \mt^2_2}{\partial \rhtb} 
L^3_2(\mt^2_2)
\right\rbrace
\nonumber \\
&+v_3 \left\lbrace
\frac{\partial^3 \mt^2_1}{\partial \rhta^2 \partial \rhtb}
L^3_1(\mt^2_1)
+\frac{\partial^3 \mt^2_2}{\partial \rhta^2 \partial \rhtb}
L^3_1(\mt^2_2)
\right\rbrace.
\label{afive} \eeq
At the minimum the mass eigenvalues of eqs. (\ref{atwo}) and their
derivatives are given by 
\beq
\mt^2_{1,2} = &( \lt \pm \gt) \kx
\nonumber \\
\left. \frac{\partial \mt^2_{1,2}}{\partial \rhta} 
\right|_{\kx} = 
\left. \frac{\partial^2 \mt^2_{1,2}}{\partial \rhta^2} 
\right|_{\kx} = &\frac{1}{2}
\left[
3 \lt + \gt + (\nua+\nub) \kx \pm 
\left. \frac{\partial A}{\partial \rhta} \right|_{\kx}
\right]
\nonumber \\
\left. \frac{\partial^2 \mt^2_{1,2}}{\partial \rhta^2} 
\right|_{\kx} = &\frac{1}{2}
\left(
5 \nua+\nub \pm 
\left.  \frac{\partial^2 A}{\partial \rhta^2} \right|_{\kx}
\right)
\nonumber \\
\left. \frac{\partial^2 \mt^2_{1,2}}{\partial \rhta \partial \rhtb} 
\right|_{\kx} = &\frac{1}{2}
\left(
6\nub \pm 
\left. \frac{\partial^2 A}{\partial \rhta \partial \rhtb} \right|_{\kx}
\right)
\nonumber \\
\left. \frac{\partial^3 \mt^2_{1,2}}{\partial \rhta^3} 
\right|_{\kx} = & \pm \frac{1}{2}
\left.  \frac{\partial^3 A}{\partial \rhta^3} \right|_{\kx}
\nonumber \\
\left. \frac{\partial^3 \mt^2_{1,2}}{\partial \rhta^2 \partial \rhtb} 
\right|_{\kx} = &\pm \frac{1}{2}
\left. \frac{\partial^3 A}{\partial \rhta^2 \partial \rhtb} \right|_{\kx},
\label{asix} \eeq
with 
\beq
A= &~2 \gt \kx
\nonumber \\
\left. \frac{\partial A}{\partial \rhta} \right|_{\kx}
= &\frac{1}{2A} \left( 8 \gt \nub \kx^2 + 8 \gt^2 \kx \right)
\nonumber \\
\left. \frac{\partial^2 A}{\partial \rhta^2} \right|_{\kx}
= &-\frac{1}{A} 
\left(
\left. \frac{\partial A}{\partial \rhta} \right|_{\kx}
\right)^2
+\frac{1}{2A} B
\nonumber \\
B = &2 \left[ 3 \lt-\gt +  (\nua-\nub) \kx \right]^2
+ 8 \nub^2 \kx^2 + 32 \gt \nub \kx 
\nonumber \\
\left. \frac{\partial^2 A}{\partial \rhta \partial \rhtb} \right|_{\kx}
= &-\frac{1}{A} 
\left(
\left. \frac{\partial A}{\partial \rhta} \right|_{\kx}
\right)^2
+\frac{1}{2A} D
\nonumber \\
D = &-2 \left[ 3 \lt-\gt +  (\nua-\nub) \kx \right]^2
+ 8 \nub^2 \kx^2 + 32 \gt \nub \kx +16 \gt^2
\nonumber \\
\left. \frac{\partial^3 A}{\partial \rhta^3} \right|_{\kx}
=&~\frac{1}{A^2} 
\left(
\left. \frac{\partial A}{\partial \rhta} \right|_{\kx}
\right)^3
-\frac{2}{A} \left(
\left. \frac{\partial A}{\partial \rhta} \right|_{\kx} 
\right)
\left(
\left. \frac{\partial^2 A}{\partial \rhta^2} \right|_{\kx} 
\right)
-\frac{1}{2 A^2} 
\left(
\left. \frac{\partial A}{\partial \rhta} \right|_{\kx}
\right) B 
+\frac{1}{2 A} B_1 
\nonumber \\
B_1 = &6(5 \nua - \nub) \left[ 3 \lt - \gt +(\nua-\nub) \kx \right]
+ 48 \nub^2 \kx
\nonumber \\
\left. \frac{\partial^3 A}{\partial \rhta^2 \partial \rhtb} \right|_{\kx}
=&~\frac{1}{A^2} 
\left(
\left. \frac{\partial A}{\partial \rhta} \right|_{\kx}
\right)^3
-\frac{2}{A} \left(
\left. \frac{\partial A}{\partial \rhta} \right|_{\kx} 
\right)
\left(
\left. \frac{\partial^2 A}{\partial \rhta \partial \rhtb} \right|_{\kx} 
\right)
-\frac{1}{2 A^2} 
\left(
\left. \frac{\partial A}{\partial \rhta} \right|_{\kx}
\right) B 
+\frac{1}{2 A} B_2 
\nonumber \\
B_2 = &-2(5 \nua - \nub) \left[ 3 \lt - \gt +(\nua-\nub) \kx \right]
+ 48 \nub^2 \kx + 64 \gt \nub.
\label{aseven} \eeq

\newpage

\section*{Figures}

\renewcommand{\labelenumi}{Fig. \arabic{enumi}}
\begin{enumerate}
\item  
: The trajectory on the critical surface 
which leads from the 
Ising to the  Heisenberg fixed point. 
\vspace{6mm}
\item  
: The $\beta$-functions for the various couplings along the
trajectory of fig. 1. 
\vspace{6mm}
\item  
: The three smallest eigenvalues for perturbations around the 
trajectory of fig. 1.
\vspace{6mm}
\item  
: The difference $\beta-\beta_I$ and the renormalized parameter
$x_R$ as a function of the value of $x$ at the cutoff. 
Lines (a) correspond to theories with renormalized mass 
$m_R/\Lx \sim 10^{-4}$ -- $10^{-3}$, while lines (b) 
correspond to theories with renormalized mass
$m_R/\Lx \sim 10^{-5}$ -- $10^{-4}$.

\end{enumerate}

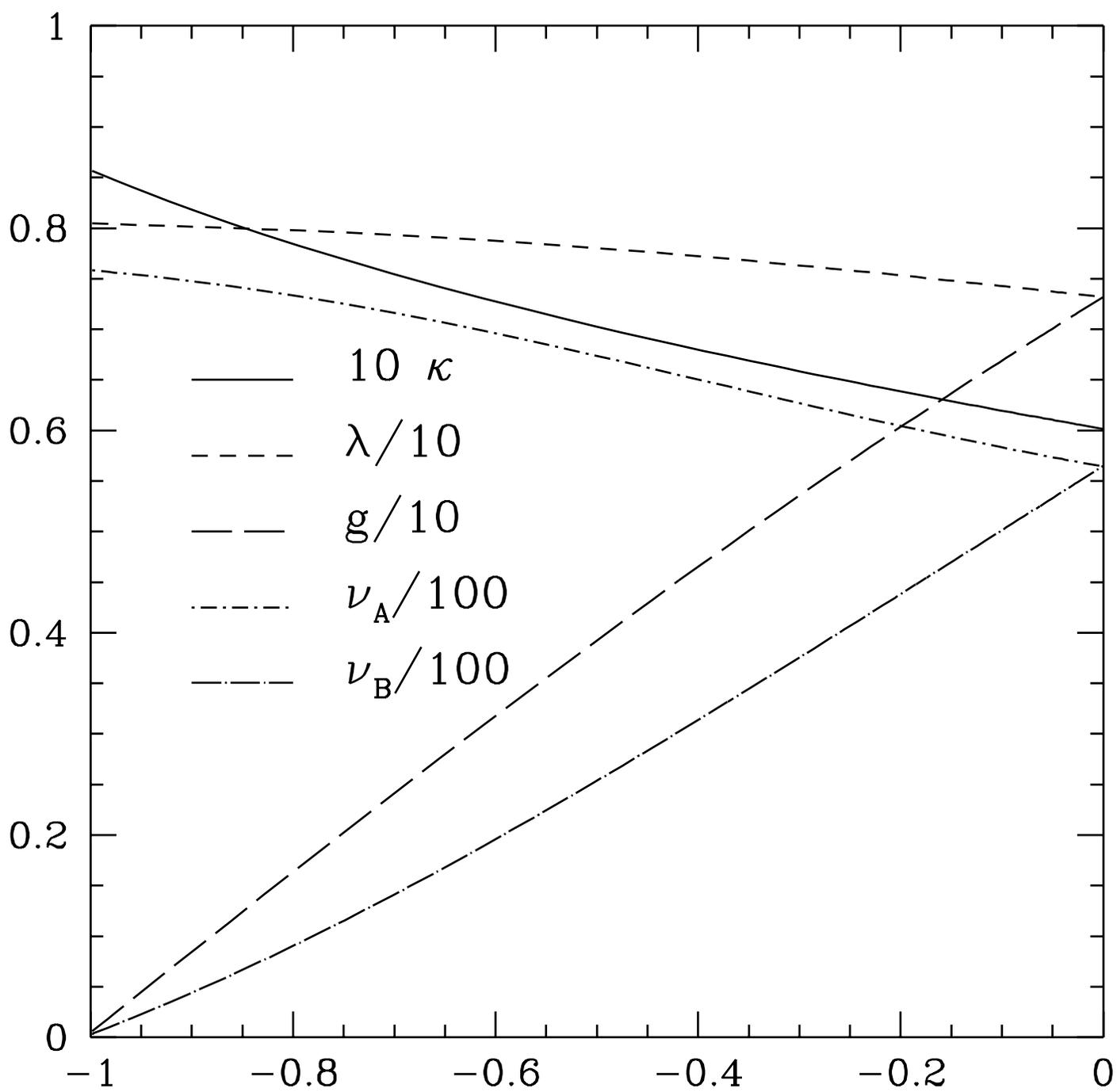
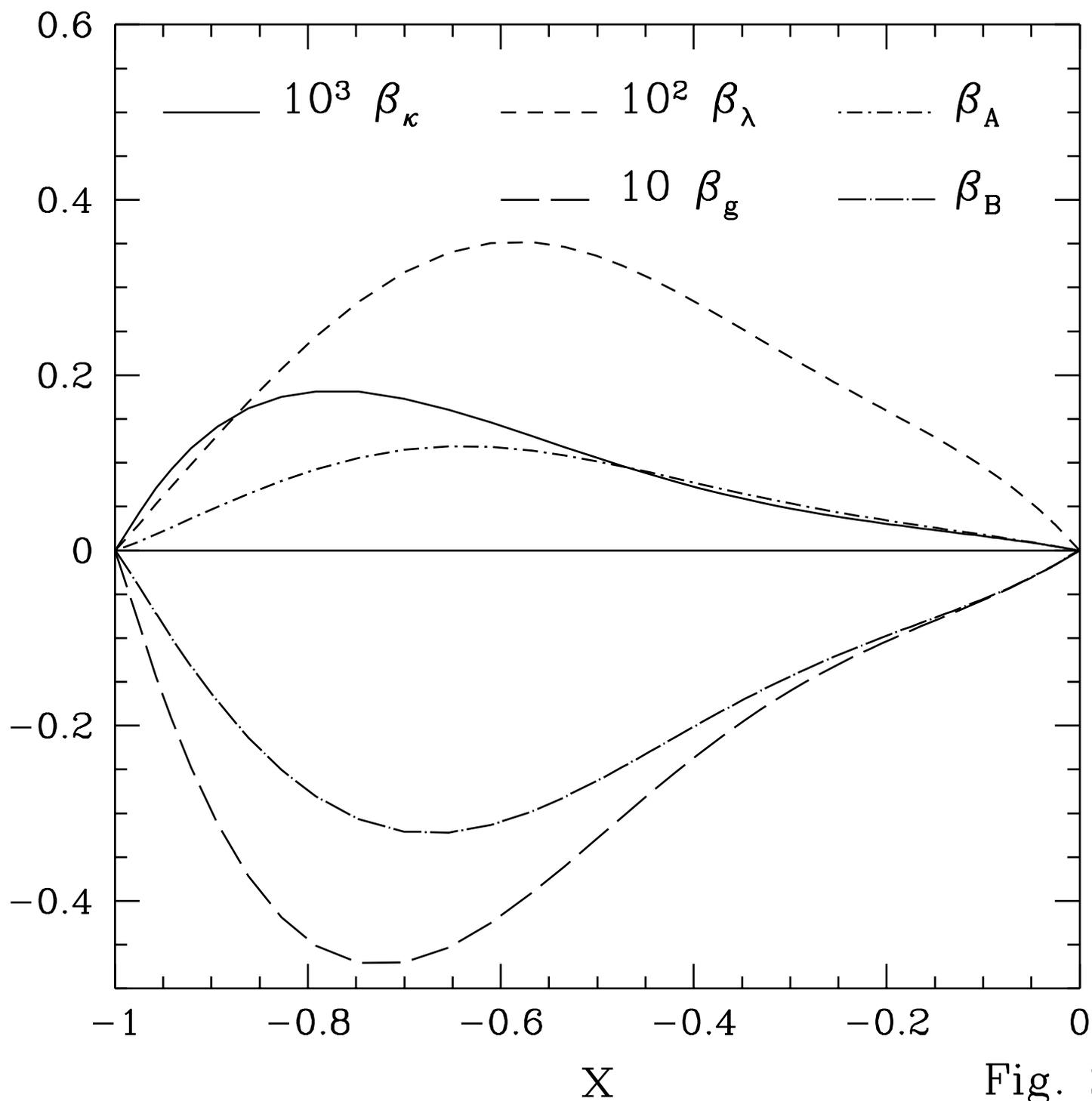
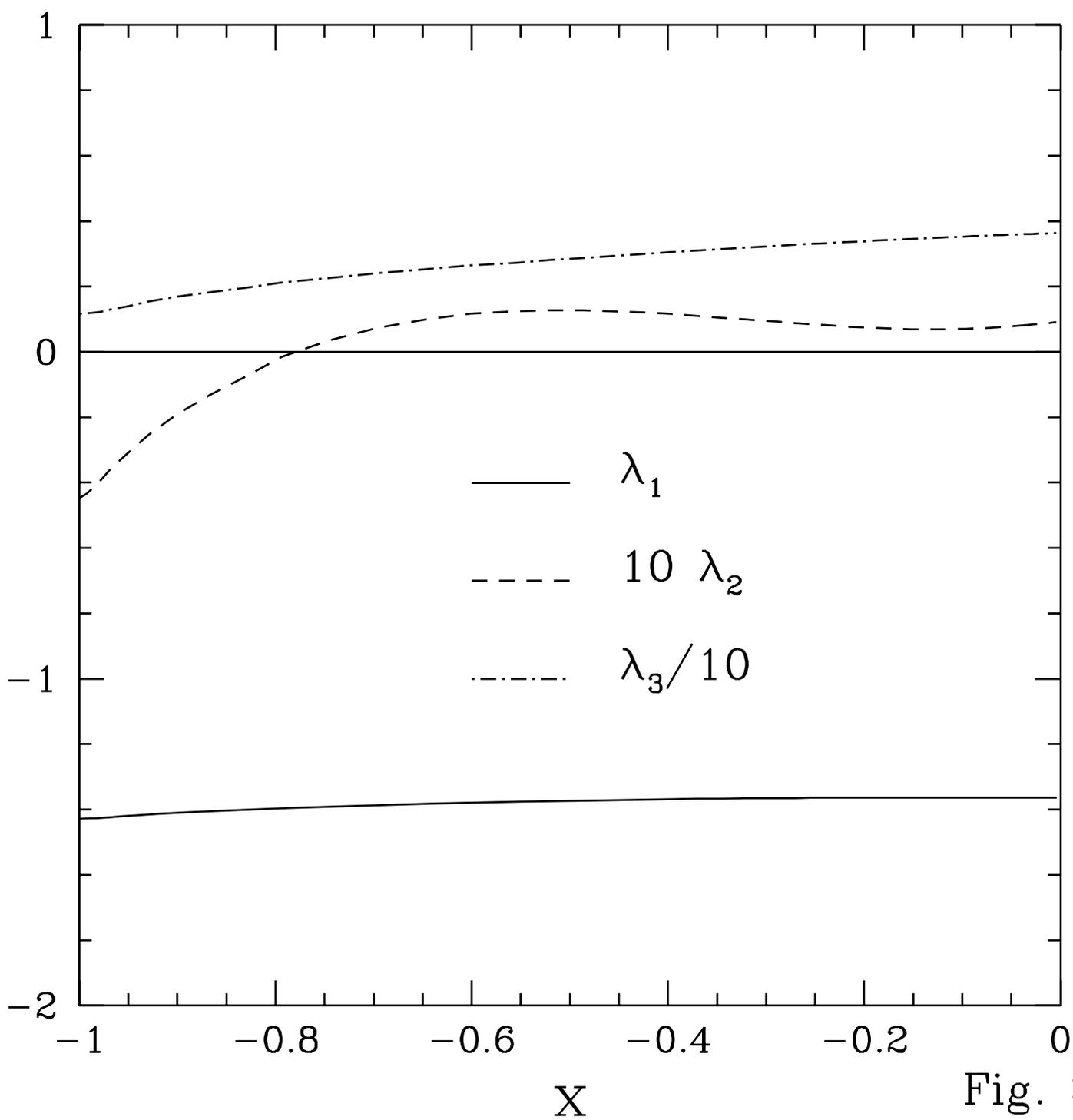
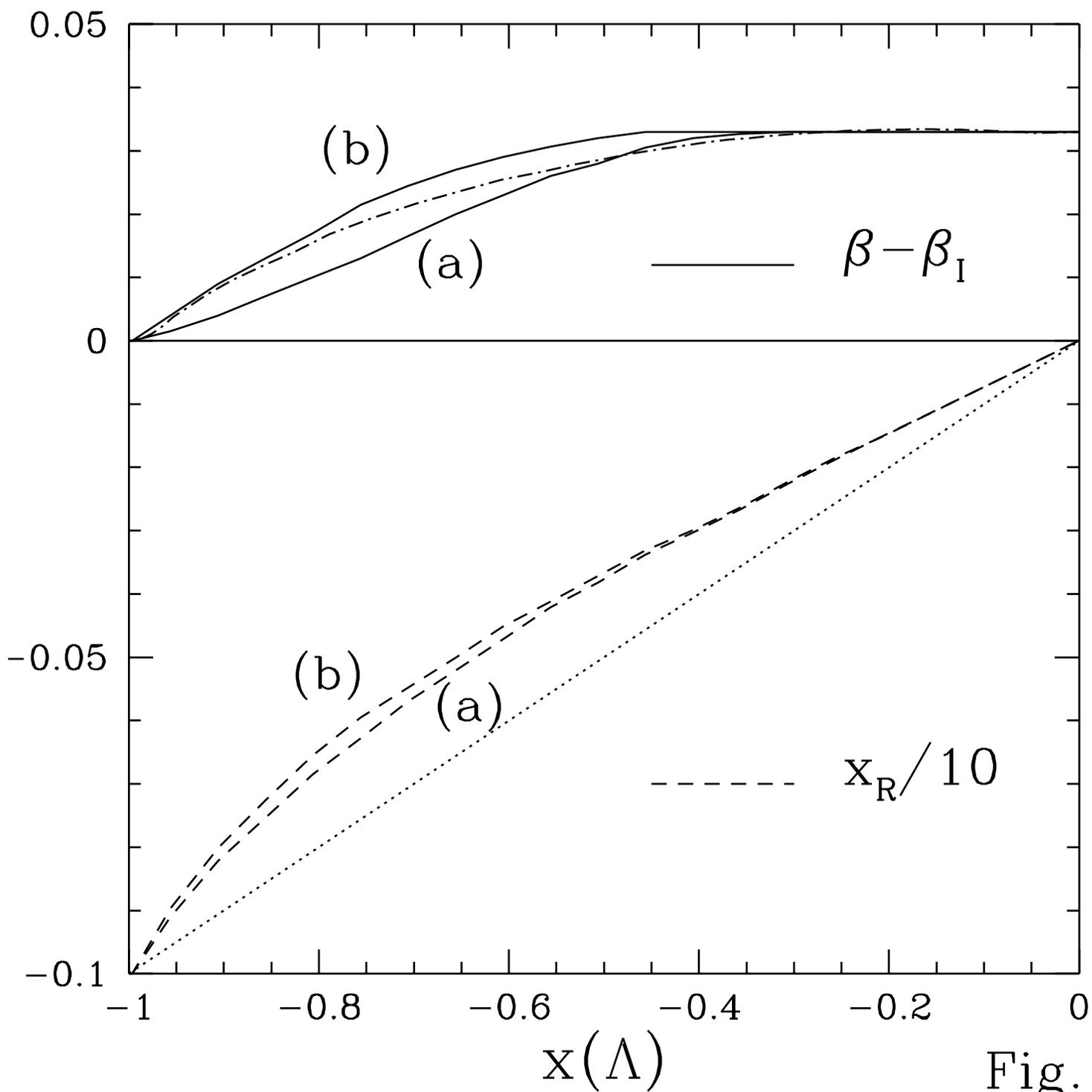
\end{document}